\documentclass[aps,twocolumn,epsfig,showpacs]{revtex4}

\usepackage{epsfig}
\usepackage{amsmath}

\begin{document}

\title{\textbf{Two Anderson impurities in a 2D host with Rashba spin-orbit interaction}}
\author{T. I. Ivanov}
\affiliation{Department of Physics, University of Sofia, 5 J. Bourchier Blvd., 1164 Sofia,
Bulgaria}
\received{\today}

\begin{abstract}
We have studied the two-dimensional two-impurity Anderson model with additional Rashba spin-orbit interaction by means of the modified perturbation theory. The impurity Green's functions we have constructed  exactly reproduce the first four spectral moments. We discuss the height and the width of the even/odd Kondo peaks as functions of the inter-impurity distance and the Rashba energy $E_R$ (the strength of the Rashba spin-orbit interaction). For small impurity separations the Kondo temperature shows a non-monotonic dependence on $E_R$ being different in the even and the odd channel. We predict that the Kondo temperature has only almost linear dependence on $E_R$ and not an exponential increase with $E_R$.

\end{abstract}
\maketitle

\section{INTRODUCTION}
Recent advancement in spintronics \cite{focus,fabian} results from an increasing interest in studying the spin-polarized transport through nanostructures \cite{awschalom}. Significant part of the research effort has been focused on the physical consequences of the spin-orbit interaction (SOI) and, in particular, on its effect in electron transport. SOI is a relativistic effect which is manifested in a coupling between electron's orbital motion and its spin \cite{winkler}. There are various proposals for spintronic devices in which the SOI plays in important role \cite{fabian}.

Kondo effect is one of the best-studied examples of correlation-driven phenomena in condensed matter systems. It consists in the screening of the magnetic moment of a localized impurity by the Fermi sea of itinerant electrons \cite{hewson}. Technological advancement allowed the Kondo effect to be studied experimentally in quantum dots \cite{gordon} and in magnetic atoms placed on a metallic surface \cite{li}. In these systems the delocalized electrons are confined to move in two dimensions. The inversion symmetry is broken by the confining potential and the so-called Rashba spin-orbit interaction (RSOI) gives rise to a coupling between electron's spin and momentum \cite{rashba,winkler}. Also, recently, investigation of magnetic impurities on graphene has been initiated \cite{brar}. Thus, a problem is posed: what is the influence of the RSOI on the Kondo effect in 2D system? Several studies have addressed it and the work has been centered on the Kondo model. It is obtained from the generic Anderson model by a suitable transformation which integrates out the charge degrees of freedom for sufficiently large Coulomb interaction between two electrons on the impurity \cite{hewson}. In Ref. 10 an effective two-channel Kondo model is derived and is predicted that the RSOI does not change the Kondo temperature except for some band-width effects. In a later work \cite{zarea}, an effective two-channel Kondo model with additional Dzyaloshinsky-Moriya term is derived. It is argued that the latter interaction produces an exponential increase of the Kondo temperature. Recently, \u{Z}itko and Bon\u{c}a \cite{zitko} on the basis of an effective single-band impurity model concluded that RSOI leads to only small variation of the Kondo temperature with the strength of the RSOI. Thus, the behavior of the 2D single-impurity Anderson (or Kondo) model with RSOI is not yet fully understood.

On the other hand, significantly less work has been done on the two-impurity Anderson model in a 2D host with RSOI. The two-impurity Anderson model (TIAM) is the simplest model that describes the interplay between the electron correlations and electron coherence which is the basic physics to be inferred from it. There are different regimes in the model due to the competition between the direct Kondo interaction and the indirect exchange (the so-called RKKY interaction). The former gives rise to the single-impurity Kondo effect while the latter is responsible for the relative orientation of the impurities' spins. Various approaches have been applied to study the TIAM - renormalization group approach \cite{krishna}, numerical computations \cite{fye}, perturbative method \cite{santoro}, and modified perturbation theory \cite{ivanov}. Recently, study of the influence of the spin-orbit interaction (of both Rashba and Dresselhaus type) on the direct Kondo and RKKY interactions has been performed in the framework of the two-impurity Kondo model \cite{mross}.

In the present work we study the two-impurity Anderson model in a 2D host with RSOI. We use the so-called modified perturbation theory which has been successfully applied to both the single-impurity \cite{martin} and the two-impurity Anderson model \cite{ivanov}. In this method an approximate electron interaction self-energy is constructed which interpolates between the limits of strong and weak coupling to the band electrons coinciding with the exact expressions in both limits. More specifically, we construct electron Green's functions that are exact up to order $\omega^{-4}$. This approach is not fully exact but gives very good results for only moderate numerical work.

The paper is organized as follows. Section II introduces the two-impurity Anderson model in a 2D system with Rashba spin-orbit interaction. The host part of the Hamiltonian is diagonalized and the  corresponding quasiparticle energies are obtained. Section III is devoted to the construction of the impurity Green's functions in the framework of the modified perturbation theory and in Section IV the results of the paper are summarized.

\section{THE MODEL HAMILTONIAN}
The Hamiltonian of two Anderson impurities, coupled to a 2D system with RSOI, is given by $H=H_{TIAM}+H_{RSOI}$ with $H_{TIAM}$ being the Hamiltonian of the two-impurity Anderson model
\begin{eqnarray}
\nonumber H_{TIAM} &=& \sum \limits_{i,{\bf k}} \epsilon_i c^+_{i \sigma} c_{i \sigma} + \sum \limits_{i} U_i n_{i\uparrow} n_{i\downarrow} +\sum \limits_{{\bf k},\sigma} \epsilon_{k\sigma} a^+_{{\bf k}\sigma} a_{{\bf k}\sigma} \\
   &+& \sum \limits_{i,{\bf k},\sigma} (V_{{\bf k}} e^{i{\bf k}.{\bf R}_i} c^+_{i \sigma} a_{{\bf k}\sigma} + H.c.).
\label{eq1}
\end{eqnarray}
Here, $c_{i \sigma}$ is the annihilation operator for an electron with spin $\sigma=\uparrow,\downarrow
$ and energy $\epsilon_i$ residing on an impurity $i=1,2$. $a_{{\bf k}\sigma}$ is the annihilation operator for a band electron with wave vector ${\bf k}$ and energy $\epsilon_{k}$. $U_i$ is the Coulomb repulsion energy between two electrons with opposite spins simultaneously occupying given impurity and $V$ is the coupling between the impurities' and band electrons. The particle-number operator is $n_{i\sigma}=c^+_{i \sigma} c_{i \sigma}$. The impurities are located in positions ${\bf R}_i$ with the distance between them being $R=|{\bf R}_1-{\bf R}_2|$. Throughout the paper we consider only the case of two identical impurities - $\epsilon_1=\epsilon_2=\epsilon$ and $U_1=U_2=U$.

The Rashba spin-orbit interaction is described by the Hamiltonian
\begin{eqnarray} \label{eq2}
 \nonumber H_{RSOI} &=& \alpha_R \sum \limits_{{\bf k}} (k_y+ik_x) a^+_{{\bf k}\uparrow} a_{{\bf k}\downarrow} + H.c. \\
   &=& \alpha_R \sum \limits_{{\bf k}} (k e^{-i\varphi_k} a^+_{{\bf k}\uparrow} a_{{\bf k}\downarrow} + H.c.)
\end{eqnarray}
where $\alpha_R$ is its strength and the phase $\varphi_k$ is defined as $k_x=-k\sin \varphi_k, k_y=\cos \varphi_k$.

In the presence of the RSOI one has to introduce the angular momentum basis for the band electrons. In polar representation the band electrons operators are developed as
\begin{equation}\label{eq3}
    a_{{\bf k}\sigma} = \sqrt{\frac{2\pi}{k}} \sum \limits^{\infty}_{m=-\infty} e^{im\varphi} a^m_{k\sigma}
\end{equation}
where $m$ is the orbital magnetic quantum number. The Hamiltonian for the band electrons is given by the sum of the third term in Eq. (\ref{eq1}) and the RSOI. It is diagonalized by the canonical transformation
\begin{equation} \label{eq4}
    a^{m+1/2}_{kh}=\frac{1}{\sqrt{2}} (a^m_{k\uparrow} + h a^{m+1}_{k\downarrow})
\end{equation}
which introduces the chirality quantum number $h=\pm 1$. Next, we switch to the even/odd basis states for the impurity electrons (in the case of identical impurities) $c_{e/o\sigma}=(c_{1\sigma} \pm c_{2\sigma})/\sqrt{2}$. With all these transformations the Hamiltonian of the system becomes $H=H_0+H_{hybr}$. $H_0$ is the Hamiltonian of the decoupled impurities' and band electrons
\begin{eqnarray} \label{eq5}
 \nonumber H_0&=& \epsilon \sum \limits_{p\sigma} n_{p\sigma} +\frac{U}{2}(n_{e\uparrow}+n_{o\uparrow})(n_{e\downarrow} + n_{o\downarrow}) \\
   &+& \sum \limits_{mh} \int \limits^{\infty}_0 \epsilon_{kh} \left (a^{m+1/2}_{kh} \right )^+ a^{m+1/2}_{kh} {\rm d}k.
\end{eqnarray}
Here, $p=e/o$ and $\epsilon_{kh}=\epsilon_{k} +\alpha_R hk$ are the renormalized band energies. The term describing the hybridization between the impurities and the band electrons is cast into the form
\begin{eqnarray} \label{eq6}
 \nonumber H_{hybr} &=& \sum \limits_{mh} \frac{1}{\sqrt{4\pi}} \int \limits^{\infty}_0 V_k \sqrt{k} J_m \left (\frac{kR}{2} \right )  [ \\
 \nonumber & &  \beta_{em}(c^+_{e\uparrow}a^{m+1/2}_{kh}+(-1)^{\frac{h-1}{2}}c^+_{e\downarrow}a^{m-1/2}_{kh})  \\
   & +&  \beta_{om}(c^+_{o\uparrow}a^{m+1/2}_{kh}+(-1)^{\frac{h-1}{2}}c^+_{o\downarrow}a^{m-1/2}_{kh} ) ]
\end{eqnarray}
where $\beta_{e/om}= (1 \pm (-1)^m)/\sqrt{2}$ and $J_m(z)$ is the $m^{{\rm th}}$ order Bessel function\cite{smetka1}. Let us point out that unlike the case of a single Anderson impurity the impurities' states couple to all orbital channels for the band electrons. In the former case, the impurity states couple only to the $m=0$ channel.

\section{THE MODIFIED PERTURBATION THEORY}
In order to discuss the dynamics of the impurities' electrons we introduce the retarded Green's functions $G_{p\sigma}(t)=-i\theta(t) \left < \{c_{p\sigma}(0),c^+_{p\sigma}(t)\} \right >$ (the curly brackets denote the anticommutator). Note that the impurity Green's functions are diagonal in the even/odd basis. The Fourier transforms of the retarded Green's functions define the interaction self-energies $\Sigma^{int}_{p\sigma}(\omega)$ for the impurities' electrons
\begin{equation}\label{eq7}
    G_{p\sigma}(\omega)=[\omega-\epsilon-\Sigma^{(0)}_{p\sigma}(\omega)-\Sigma^{int}_{p\sigma}(\omega)]^{-1}.
\end{equation}
$\Sigma^{(0)}_{p\sigma}(\omega)$ is the Hartree-Fock elastic self-energy \cite{smetka2}
\begin{equation}\label{eq8}
 \Sigma^{(0)}_{p\sigma}(\omega)=\sum \limits_{h}\int \limits^{\infty}_0 V^2_k \frac{1 \pm J_0(kR)}{\omega-\epsilon_{kh}+i0^+}\frac{k {\rm d}k}{4\pi}.
\end{equation}

To proceed with further discussion of the physics of the system we are studying we must make some kind of approximation for the interaction self-energies. In this work, we choose to employ the so-called modified perturbation theory. In it the interaction self-energy is approximated by the ansatz
\begin{equation}\label{eq9}
    \Sigma^{int}_{p\sigma}(\omega)=\frac{U}{2}\left <n_{e-\sigma}+n_{o-\sigma} \right > + \frac{A_{\sigma} \Sigma^{(2)}_{p\sigma}(\omega)}{1-B_{\sigma} \Sigma^{(2)}_{p\sigma}(\omega)}.
\end{equation}
The first term is the contribution of the Hartree-Fock decoupling of the Coulomb interaction term, $\Sigma^{(2)}_{p\sigma}(\omega)$ is the second-order self-energy which is obtained in a perturbation theory with respect to the Coulomb repulsion energy $U$, and the coefficients $A_{\sigma}$ and $B_{\sigma}$ are to be determined. The explicit expression for $\Sigma^{(2)}_{p\sigma}(\omega)$ is as follows
\begin{eqnarray} \label{eq10}
\nonumber \Sigma^{(2)}_{e/o\sigma}(\omega) &=& \frac{U^2}{4} \int \limits^{\infty}_{-\infty} \frac{{\rm d}\omega_1{\rm d}\omega_2{\rm d}\omega_3}{\omega+\omega_1-\omega_2-\omega_3+i0^+} \\
\nonumber &\times& F(\omega_1,\omega_2,\omega_3)   [S_1(\omega_1,\omega_2) \rho^{(0)}_{e/o}(\omega_3) \\
 &+& S_2(\omega_1,\omega_2) \rho^{(0)}_{o/e}(\omega_3)].
\end{eqnarray}
The auxiliary quantities are defined as $F(\omega_1,\omega_2,\omega_3)=[1-f(\omega_1)]f(\omega_2)f(\omega_3)+f(\omega_1)[1-f(\omega_2)][1-f(\omega_3)]$, $S_1(\omega_1,\omega_2)=\rho^{(0)}_{e}(\omega_1)\rho^{(0)}_{e}(\omega_2)+\rho^{(0)}_{o}(\omega_1)\rho^{(0)}_{o}(\omega_2)$, and $S_2(\omega_1,\omega_2)=\rho^{(0)}_{e}(\omega_1)\rho^{(0)}_{o}(\omega_2)+\rho^{(0)}_{o}(\omega_1)\rho^{(0)}_{e}(\omega_2)$. Here, $f(\omega)$ is the Fermi-Dirac distribution function and $\rho^{(0)}_{p}(\omega)=- {\rm Im} \ G^{(0)}_{p\sigma}(\omega)$ is the Hartree-Fock spectral function. In the modified perturbation theory, the second-order self-energy is calculated using the following expression for the Hartree-Fock Green's functions
\begin{equation}\label{eq11}
    G^{(0)}_{p\sigma}(\omega)=[\omega-\tilde{\epsilon} - U \left <n_{e-\sigma}+n_{o-\sigma} \right >/2 -\Sigma^{(0)}_{p\sigma}(\omega)]^{-1},
\end{equation}
the impurity energy level $\epsilon$ being replaced by an auxiliary energy level $\tilde{\epsilon}$ which will take into account various renormalizations due to the Coulomb interaction and the charge transfer between the impurities and the bands. Later on we shall briefly discuss the ways to determine $\tilde{\epsilon}$.

The electron Green's functions given by Eq. (\ref{eq7}) with the interacting self-energy from Eq. (\ref{eq9}) may be constructed in such a way that the first four moments of the corresponding spectral functions are exactly reproduced\cite{potthoff}, that is, the $G_{p\sigma}(\omega)$ we are going to obtain will be exact up to order $\omega^{-4}$. To that goal, we develop the exact Green's function $G_{p\sigma}(\omega)$ [Eq. (\ref{eq7})] in series with respect to $1/\omega$
\begin{equation}\label{eq12}
G_{p\sigma}(\omega)=\sum \limits^{\infty}_{n=0} \frac{M^{(n)}_{p\sigma}}{\omega^{n+1}}.
\end{equation}
The exact expression for the moment $M^{(n)}_{p\sigma}$ is given by
\begin{equation}\label{eq12p}
    M^{(n)}_{p\sigma}=\left < \{\hat{L}^n c_{p\sigma},c^+_{p\sigma}\} \right >
\end{equation}
where the operator $\hat{L}$ acts as $\hat{L} c_{p\sigma} =[c_{p\sigma},H]$ (square brackets denote commutator)\cite{potthoff}. Next, we develop the approximate Green's function $G_{p\sigma}(\omega)$ given by Eqs. (\ref{eq7},\ref{eq9},\ref{eq10}) in series with respect to $1/\omega$. This is achieved by developing the explicit expressions for the elastic self-energy $\Sigma^{(0)}_{p\sigma}(\omega)$ [Eq. (\ref{eq8})] and the second-order self-energy $\Sigma^{(2)}_{p\sigma}(\omega)$ [Eq. (\ref{eq10})] in series with respect to $1/\omega$. We are able to determine the coefficients $A_{\sigma}$ and $B_{\sigma}$ in Eq. (\ref{eq9}) in such a way that the first four terms in Eq. (\ref{eq12}) coincide with the first four terms in the expansion of the approximate Green's function, that is the approximate Green's function reproduces exactly the first four moments $M^{(n)}_{p\sigma}, n=0,1,2,3$. Let us point out that the second-order perturbation theory with respect to $U$ (it corresponds to $A_{\sigma}=1$ and $B_{\sigma}=0$) reproduces only the first two moments. We shall not give details of the calculations because they are too cumbersome to be presented here. The calculations along these lines but in the case of the infinite-dimensional Hubbard model can be found in Ref. 21. The explicit expressions for $A_{\sigma}$ and $B_{\sigma}$ can be cast into the form
\begin{equation}\label{eq13}
    A_{\sigma}=\frac{(n_{e\sigma}+n_{o\sigma})(2-n_{e\sigma}-n_{o\sigma})}{(n^{(0)}_{e\sigma}+n^{(0)}_{o\sigma})(2-n^{(0)}_{e\sigma}-n^{(0)}_{o\sigma})},
\end{equation}
\begin{equation}\label{eq14}
   B_{\sigma}=\frac{b_{\sigma}-b^{(0)}_{\sigma}+U(1-n_{e\sigma}-n_{o\sigma})}{U^2(n^{(0)}_{e\sigma}+n^{(0)}_{o\sigma})(2-n^{(0)}_{e\sigma}-n^{(0)}_{o\sigma})/4}
\end{equation}
where $n_{p\sigma}$ is the average number of impurity electrons in the corresponding channel $p$. It is calculated self-consistently by solving the equations
\begin{equation}\label{eq15}
   n_{p\sigma}=-\int \limits^{\infty}_{-\infty} \frac{{\rm d}\omega}{\pi}f(\omega) {\rm Im} \ G_{p\sigma}(\omega).
\end{equation}
$n^{(0)}_{p\sigma}$ are auxiliary numbers of particles defined as
\begin{equation}\label{eq16}
      n^{(0)}_{p\sigma}=-\int \limits^{\infty}_{-\infty} \frac{{\rm d}\omega}{\pi}f(\omega) {\rm Im} \ G^{(0)}_{p\sigma}(\omega)
\end{equation}
with $G^{(0)}_{p\sigma}(\omega)$ from Eq. (\ref{eq11}).
\begin{equation}\label{eq17}
  b_{\sigma}=\epsilon+\frac{4\sum\limits_{{\bf k}} V_k \left < a^+_{{\bf k}-\sigma}c_{1 -\sigma}(2n_{1-\sigma}-1)\right> e^{i{\bf k}.{\bf R}_1}}{U^2(n_{e\sigma}+n_{o\sigma})(2-n_{e\sigma}-n_{o\sigma})},
\end{equation}
\begin{equation}\label{eq18}
  b^{(0)}_{\sigma}=\tilde{\epsilon}+\frac{4(n^{(0)}_{e-\sigma}+n^{(0)}_{o-\sigma}-1)\sum\limits_{{\bf k}} V_k \left < a^+_{{\bf k}-\sigma}c_{1 -\sigma}\right>^{(0)} e^{i{\bf k}.{\bf R}_1}}{U^2(n^{(0)}_{e\sigma}+n^{(0)}_{o\sigma})(2-n^{(0)}_{e\sigma}-n^{(0)}_{o\sigma})}.
\end{equation}
The superscript $(0)$ in the correlation function in Eq. (\ref{eq18}) means that it is calculated at the Hartee-Fock level of approximation. The correlation functions in Eqs. (\ref{eq17}), (\ref{eq18}) are obtained using the equation-of-motion method and the result is
\begin{eqnarray}\label{eq19}
\nonumber &\sum\limits_{{\bf k}}& V_k \left < a^+_{{\bf k}-\sigma}c_{1 -\sigma}(2n_{1-\sigma}-1)\right> e^{i{\bf k}.{\bf R}_1} = -{\rm Im} \ \int \limits^{\infty}_{-\infty} \frac{{\rm d}\omega}{\pi}  \\
   &\times& f(\omega) \sum\limits_{p=e,o}\Sigma^{(0)}_{p-\sigma}(\omega) \left( \frac{2\Sigma_{p-\sigma}(\omega)}{U} -1 \right)G_{p-\sigma}(\omega),
\end{eqnarray}
\begin{eqnarray}\label{eq20}
 \nonumber &\sum\limits_{{\bf k}}& V_k \left < a^+_{{\bf k}-\sigma}c_{1 -\sigma}\right>^{(0)} e^{i{\bf k}.{\bf R}_1}=-{\rm Im} \ \int \limits^{\infty}_{-\infty} \frac{{\rm d}\omega}{\pi}f(\omega) \\
   &\times& \sum\limits_{p=e,o} \Sigma^{(0)}_{p-\sigma}(\omega) G^{(0)}_{p-\sigma}(\omega).
\end{eqnarray}

The coefficient $A_{\sigma}$ [Eq. (\ref{eq13})] ensures that the high-frequency limit of the exact Green's functions [Eq. (\ref{eq7})] coincides with the high-frequency limit of the corresponding approximate expressions \cite{ivanov}. The approximate Green's functions have the correct limit of zero coupling between the impurities and the band electrons $V_k \to 0$ (this limit is exactly solvable)\cite{ivanov}. On the other hand, in the limit of strong coupling to the band electrons $U/V_k \to 0$, $G_{p\sigma}(\omega)$ obviously coincides with the corresponding exact expression. Thus, the Green's functions we have obtained interpolate between the two exact limits of strong and weak coupling to the band electrons.

In order to complete the construction of the impurity Green's functions we have to choose the parameter $\tilde{\epsilon}$. A discussion on the possible ways to fix its value in the case of the infinite-dimensional Hubbard model have shown that none of the possible choices is to be preferred over the others\cite{potthoff}. In a previous paper\cite{ivanov} on the two-impurity Anderson model (without spin-orbit interaction), we have imposed the Friedel sum rule to determine the value of $\tilde{\epsilon}$. The Friedel sum rule relates the average impurity charge and the phase shift at the Fermi level at zero temperature and represents an exact condition on the $\omega \to 0$ behavior of the Green's function. However, its application is confined only to the zero-temperature case while we intent to discuss the nonzero temperature behavior of the model under consideration. Therefore, we choose the value of $\tilde{\epsilon}$ from the condition $n_{p\sigma}=n^{(0)}_{p\sigma}$. We have verified that, indeed, there is no significant difference between the results obtained at $T=0$ with the two choices for $\tilde{\epsilon}$ we have considered. Now the modified perturbation theory for the TIAM with RSOI is completed. Let us stress that, unlike the studies of the two-impurity Kondo model, no degrees of freedom are integrated out and both the charge and the spin fluctuations are treated on equal footing in our approach.

To proceed with the presentation of our results we need to specify the dispersion of the band electrons. The band electrons have a quadratic dispersion $\epsilon_k=k^2/2m^*-E_0$ where $m^*$ is the effective mass and $E_0$ is the bottom of the band. The dispersion of the band electrons with given chirality can be cast into the form
\begin{equation}\label{eq21}
    \epsilon_{kh}=(k+hk_R)^2/2m^*-E_0-E_R
\end{equation}
with $k_R=m^* \alpha_R$ being the Rashba wave vector and $E_R=k_R^2/2m^*$ is the Rashba energy. In the presence of the Rashba spin-orbit interaction the band bottom is actually at $E_R+E_0$. The Fermi wave vector is defined as $k_F=\sqrt{2m^*E_0}$ (we choose the Fermi energy to be 0). We calculate the Hartree-Fock self-energy [Eq. (\ref{eq8})] with the dispersion $\epsilon_{kh}$ [Eq. (\ref{eq21})]. The imaginary parts can be computed analytically but in order to obtain the real parts one has to resort to numerical integration. The result for the imaginary part ${\rm Im} \ \Sigma^{(0)}_{p\sigma}(\omega)$ is too complicated to be presented here. We shall only point out that it has a characteristic dependence on the band bottom of the form $\sqrt{E_R/(\omega+E_0+E_R)}$ (see below)\cite{zitko}.

\section{RESULTS}
The numerical procedure involves the self-consistent solution of Eq. (\ref{eq15}). The impurity Green's functions depend on the values of the correlation functions in Eq.(\ref{eq19}) which in turn depend on $G_{p\sigma}(\omega)$. Therefore, another self-consistent procedure is required in order to compute the value of $G_{p\sigma}(\omega)$ which is used in Eq. (\ref{eq15}). The procedure is rather effective and only in the case of small impurity separations $k_FR < 0.75$  significant number of steps have to be performed in order to achieve self-consistency. In the following, we present our results for the set of parameters $\epsilon=-U/2$, $U=3\pi\gamma$, $E_0=0.7U$ with $\gamma=V^2m^*/4$. Note that the second-order perturbation theory in $U$ is not expected to work for such large value of $U$. The choice for the value of $\epsilon$ does not represent some special point for which a specific behavior can be expected (recall that the single-impurity Anderson model (SIAM) with $\epsilon=-U/2$ and with a symmetric density of states for the band electrons is a particle-hole symmetric model). We have pointed out previously \cite{ivanov} that for TIAM  with $\epsilon=-U/2$ one obtains different physics compared to the case of symmetric SIAM. The reason is that in the case of TIAM there are two different effective energy levels in the even/odd channel given by $\epsilon_{e/o}=\epsilon + {\rm Re} \ \Sigma^{(0)}_{e/o} (0)$ as well as two different elastic level widths $\gamma_{e/o}$ (see below). Thus, even with a symmetric density of states for the band electrons TIAM is not a particle-hole symmetric model. Moreover, in the present case, the 2D band electrons with the dispersion $\epsilon_{kh}$ [Eq. (\ref{eq21})] are not actually a particle-hole symmetric system. These considerations suggest that the behavior of the TIAM with RSOI we are going to discuss will not qualitatively depend on any specific value of $\epsilon$.

In Figs. 1 and 2, we show the spectral functions ({\it i.e.} $-{\rm Im} \ G_{p\sigma}(\omega)/\pi$) of the impurities' electrons in the even and odd channels at zero temperature $T=0$. The results are presented in Fig. 1 for Rashba energy $E_R=0.1U$ and for several impurity separations $k_FR$ and in Fig. 2 for impurity separation $k_FR=1$ and several values of $E_R$. There are features in the left-hand side of both figures (at $\omega \sim -7.5\gamma$ in Fig. 1) that result from the dependence of ${\rm Im} \ \Sigma^{(0)}_{p\sigma}(\omega)$ on the band bottom we have mentioned above. As evidenced from the figures, the height and the width of the Kondo peaks depend on both $E_R$ and $k_FR$.

The results for the dependence of the height and the width of the Kondo peaks on $E_R$ and $k_FR$ are presented in the insets of Fig. 2 and in Fig. 3. The $k_FR$ dependence shown in Fig. 3 is qualitatively similar to the dependence of the height and the width of the even/odd Kondo peaks on $k_FR$ for the 3D TIAM without RSOI \cite{ivanov}. The height of the even/odd Kondo peaks [Fig. 3, solid lines] increases/decreases with $k_FR$ for both values of $E_R$ and reaches extremum (maximum/minimum) for some value of $k_FR$. Semi-quantitative understanding of this behavior has been suggested in Ref. 16. The relevant frequency range for the Kondo effect is the range of small $|\omega|$ ($|\omega|$ less than several times the Kondo temperature) and in it one can approximate the effective even/odd elastic level widths $\gamma_p=-{\rm Im} \ \Sigma^{(0)}_{p\sigma}(0)$ by
\begin{eqnarray}\label{eq22}
  \nonumber \gamma_{e/o} = & \gamma & \left \{ \left ( 1+\sqrt{\frac{E_R}{E_0+E_R}} \right ) ( 1 \pm  j_+ ) \right.\\
   &+&  \left . \left (1-\sqrt{\frac{E_R}{E_0+E_R}}\right ) (1 \pm j_- ) \right \}
  \end{eqnarray}
where
\begin{equation}\label{eq23}
   j_{+/-}=J_0 \left( k_FR \left ( \sqrt{\frac{E_R}{E_0}} \pm \sqrt{1+\frac{E_R}{E_0}} \right ) \right ).
\end{equation}
In the case $E_R=0$ the effective level width is $\gamma_{e/o}=2\gamma(1\pm J_0 (k_FR))$. The extremums of this function are given by the zeros of $J_1 (k_FR)$ and the first zero is $k_FR=3.8317$ \cite{janke}. Thus, $\gamma_{e}$ has a minimum and $\gamma_{o}$ has a maximum at this value of $k_FR$. The data in Fig. 3 show that the height of the even/odd Kondo peaks reaches a maximum/minimum around $k_FR \approx3.8$ in agreement with the previous discussion \cite{ivanov}. When $E_R=0.1U$ the minumum/maximum of $\gamma_{e/o}$ is realized for $k_FR \approx 2.9$ and, indeed, the height of the even/odd Kondo peaks has a maximum/minimum in the vicinity of this value. Further, $J_0 (k_FR)=0$ for $k_FR=2.4048$ \cite{janke} and, consequently, for $E_R=0$ $\gamma_{e}=\gamma_{o}$ for this value of $k_FR$ and both are equal to the elastic level width in the case $k_FR \to \infty$ which corresponds to the doubly-degenerate case of two identical infinitely-separated Anderson impurities. Fig. 3 shows that the height of the even and of the odd Kondo peak is almost equal to 1 in the vicinity of $k_FR \approx 2.4$ (recall that the data are normalized to the corresponding results for $k_FR \to \infty$ {\it i.e.} the case of a single Anderson impurity).

The width of the even Kondo peak reaches maximum at $k_FR \approx 1.7$ for $E_R=0$ and at somewhat smaller value of $k_FR \approx 1.5$ for $E_R=0.1U$ [Fig. 3, lower panel, the dashed lines]. In a previous study \cite{ivanov}, we have related the occurrence of this maximum to the known change of the sign of the instantaneous spin-spin correlation function which in the case of 3D TIAM without RSOI takes place at $k_FR =3\pi/4$ \cite{santoro,fye}. We are not aware of previous works that have calculated the instantaneous spin-spin correlation function in the model we are studying. But, based on our previous experience, we can suppose that in the case of 2D TIAM with RSOI the instantaneous spin-spin correlation function changes sign in the vicinity of $k_FR \approx 1.7$ for $E_R=0$ and at $k_FR \approx 1.5$ for $E_R=0.1U$ being ferromagnetic/antiferromagnetic for lower/higher values of $k_FR$.

The width of the Kondo peak in the odd channel is a monotonically increasing function of $k_FR$ [Fig. 3, upper panel, the dashed lines]. It has a maximum at the same value of $k_FR$ at which a minimum of the height of the odd Kondo peak is realized (see above).  At the same value of $k_FR$ there is a maximum of the height and a minimum of the width of the even Kondo peak [Fig. 3]. Physically, maximum of the height and minimum of the width of the even Kondo peak means that it is most difficult for the even Kondo resonance  to be destroyed by the antiferromagnetic correlations of the impurity spins at the corresponding impurity separation. On the other hand, the odd Kondo peak has minimal height and maximal width for the same impurity separation, that is, the antiferromagnetic impurity correlations tend to most easily destroy the odd Kondo resonance.

The results for the dependence of the height and the width of the even/odd Kondo peaks on the Rashba energy $E_R$ are presented in the insets of Fig. 2. The $E_R$ dependence is qualitatively the same in both the even and the odd channel. The height has a maximum at some value of $E_R$ (different for the even and the odd channel) and the width is a monotonically increasing function of $E_R$.

Finally, we show the results for the dependence of the Kondo temperature $T_K$ on $E_R$ [Fig. 4]. The Kondo temperature has been determined from the variation of the width of the Kondo peak in the even/odd channel with the temperature. We are not aware of previous works that have considered the reliability of obtaining $T_K$ in the framework of the modified perturbation theory. Nevertheless, we think that this approach provides reasonable predictions for the dependence of $T_K$ on the model parameters. As evidenced from Fig. 4, $T_K$ has a non-monotonic dependence on $E_R$ for sufficiently small impurity separation $k_FR < 1.2 - 1.3$. For so small impurity separation there is a significant difference between the Kondo temperatures in the even and in the odd channels with $T_{Ke} > T_{Ko}$. Recall that in the limit $k_FR \to 0$ (the two impurities merge at some point) the odd channel disappears and the even channel corresponds to one impurity with elastic level width $\gamma_e (\omega)$. For larger values of $E_R > 0.1-0.12U$ the dependence of $T_K$ is almost linear. For larger impurity separations the even and the odd Kondo temperature are practically equal and their $E_R$ dependence is monotonic - $T_K$ increases almost linearly with $E_R$. Thus, we predict that the inclusion of the RSOI will not lead to an exponential increase of $T_K$. For larger values of $E_R$ the Kondo temperature increases almost linearly for any value of $R$ similarly to the result obtained in a different approach in the single-impurity Anderson model \cite{zitko}.

To conclude, we have constructed a modified perturbation theory for the 2D two-impurity Anderson model with Rashba spin-orbit interaction. The ansatz interacting self-energy is chosen in such a way that the impurity Green's functions are exact up to order $\omega^{-4}$. The height and width of the Kondo peaks in the even/odd channels are obtained numerically as functions of the inter-impurity distance and the strength of the Rashba spin-orbit interaction. It is predicted that the Kondo temperature will not have an exponential increase with the Rashba energy instead having only almost linear dependence on it.

\begin{figure}[ht]
\vspace{0.5cm} \epsfxsize=7.0cm \hspace*{-1.5cm} \epsfbox{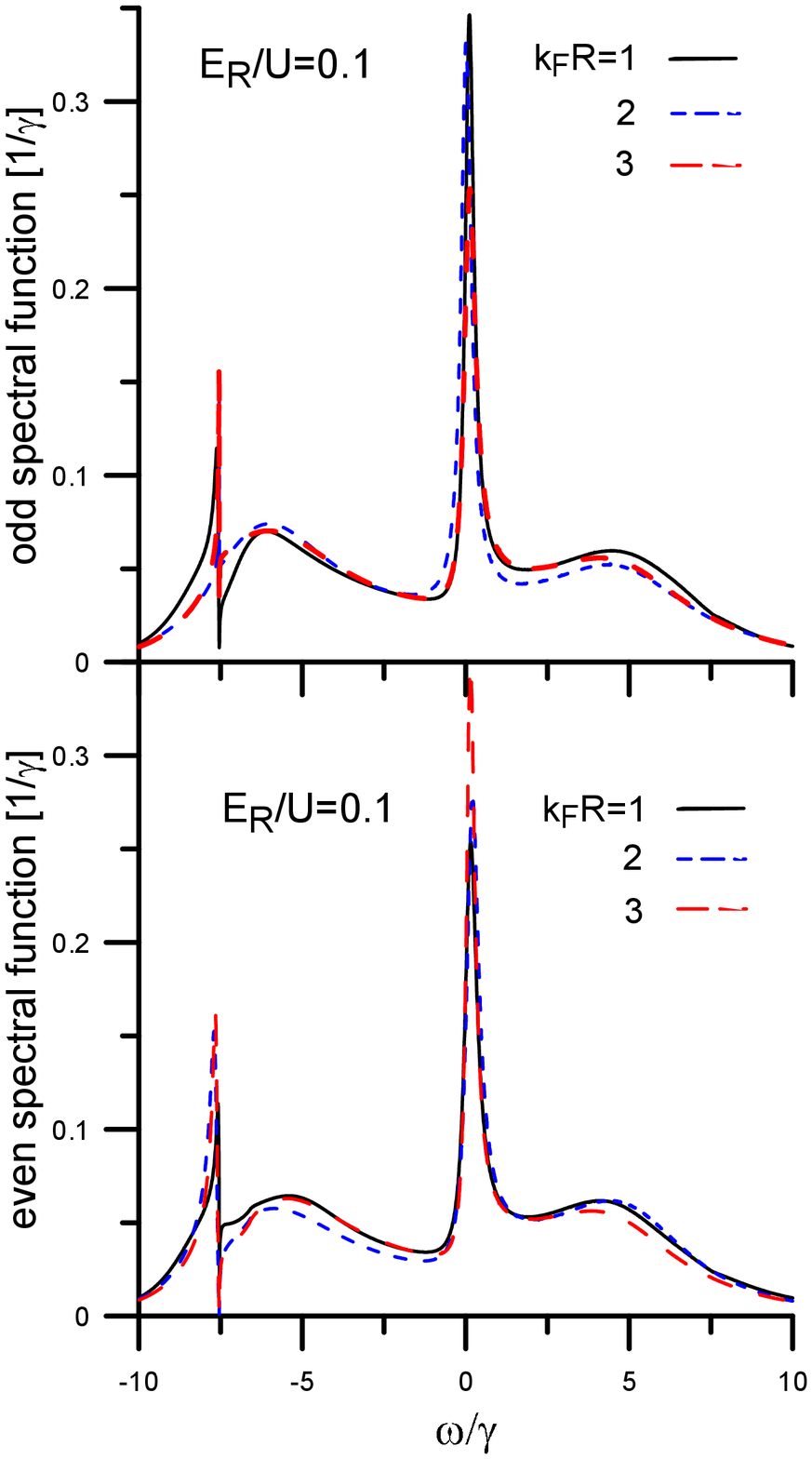}
\caption{(Color online) The impurity spectral functions in the even/odd channels for $E_R=0.1U$ and several values of $k_FR$.}
\label{fig1}
\end{figure}

\begin{figure}[ht]
\vspace{0.5cm} \epsfxsize=7.0cm \hspace*{-1.5cm} \epsfbox{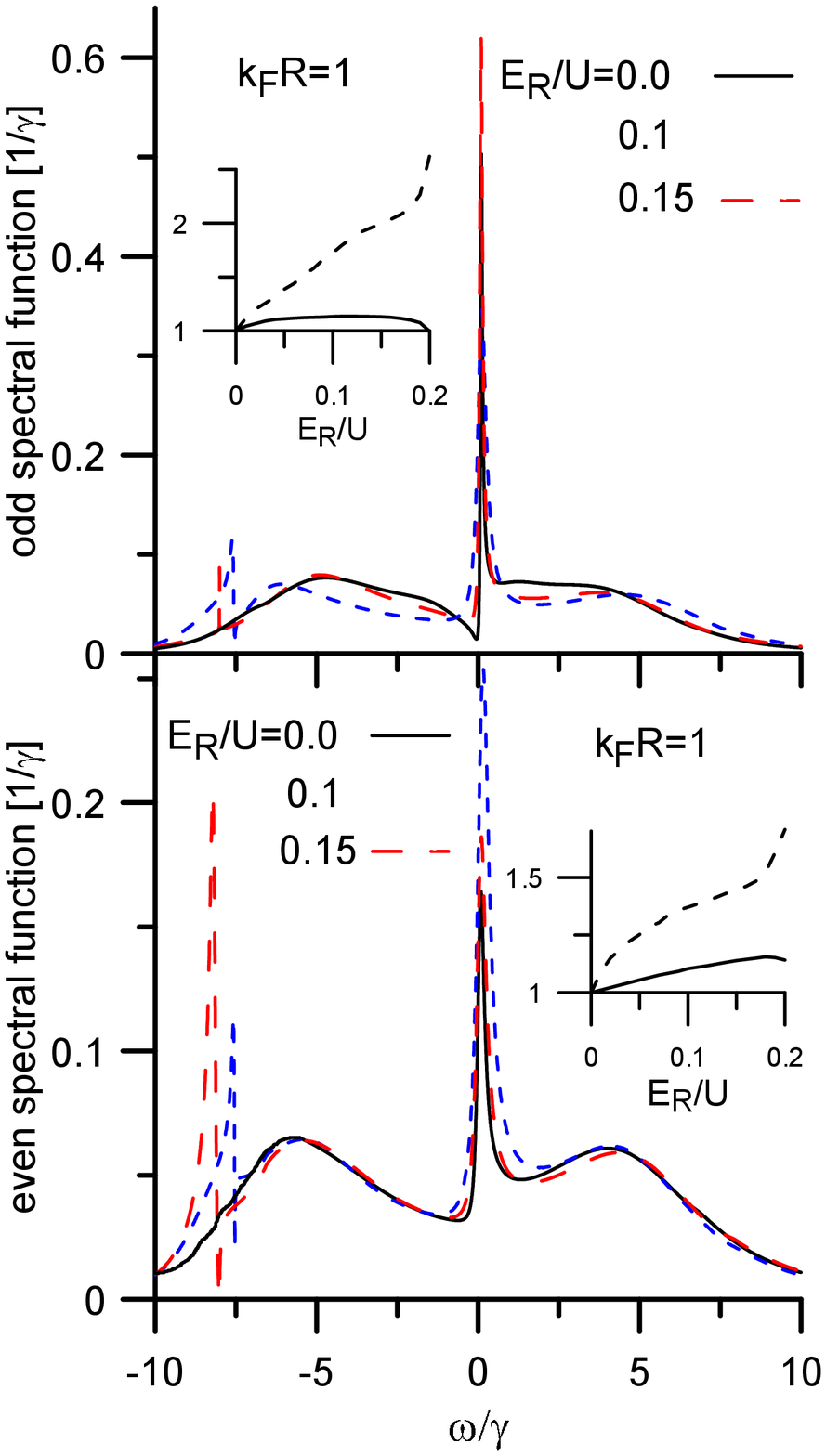}
\caption{(Color online) The impurity spectral functions in the even/odd channels for $k_FR=1$ and several values of $E_R$. The insets: The $E_R$ dependence of the height (solid line) and the width (dashed line) of the even/odd Kondo peaks.}
\label{fig2}
\end{figure}

\begin{figure}[ht]
\vspace{0.5cm} \epsfxsize=7.0cm \hspace*{-1.5cm} \epsfbox{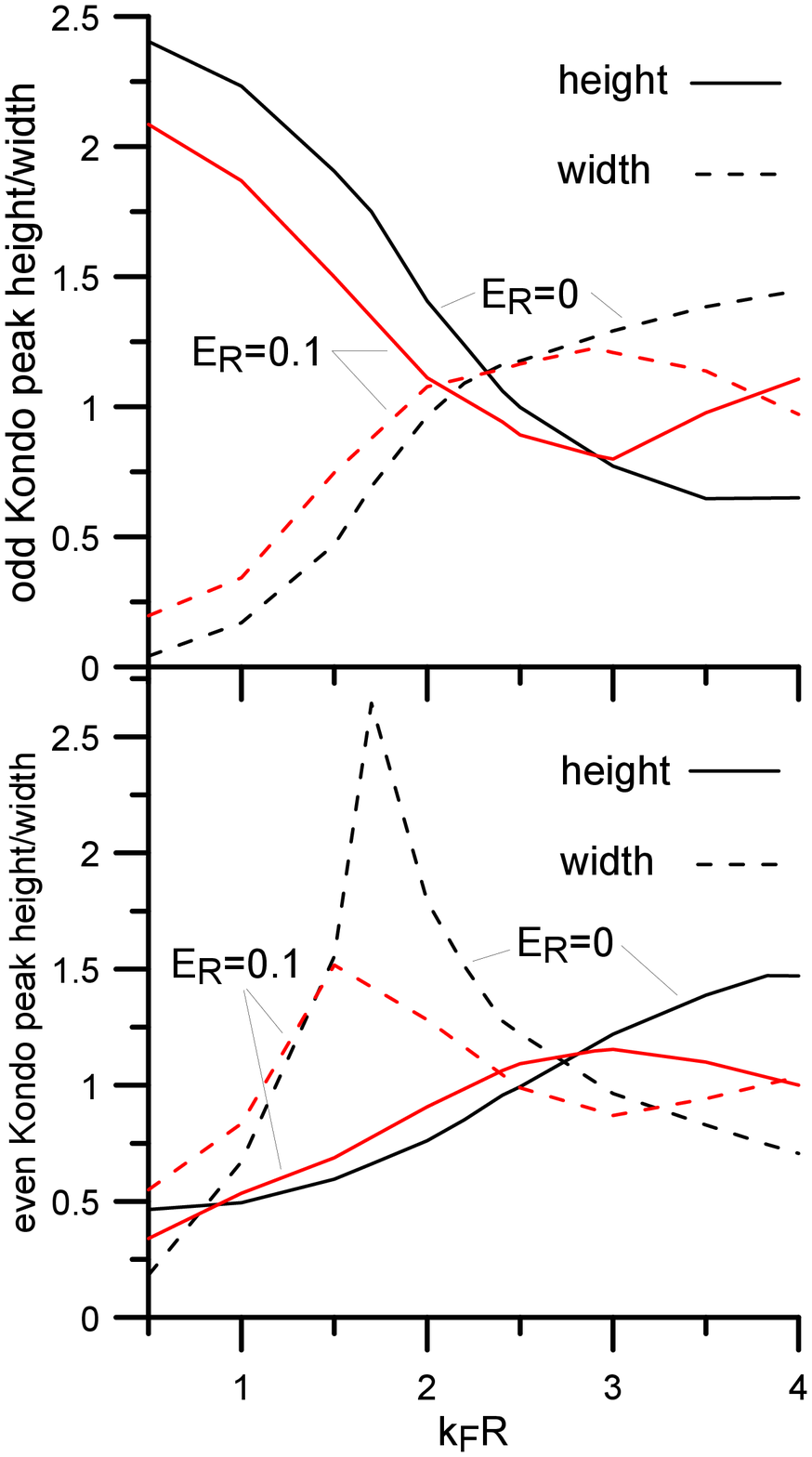}
\caption{(Color online) The $k_FR$ dependence of the height (solid line) and the width (dashed line) of the even/odd Kondo peaks for $E_R=0$ and $E_R=0.1U$. The data are normalized with respect to the corresponding quantities in the limit $k_FR \to \infty$.}
\label{fig3}
\end{figure}

\begin{figure}[ht]
\vspace{0.5cm} \epsfxsize=7.0cm \hspace*{-1.5cm} \epsfbox{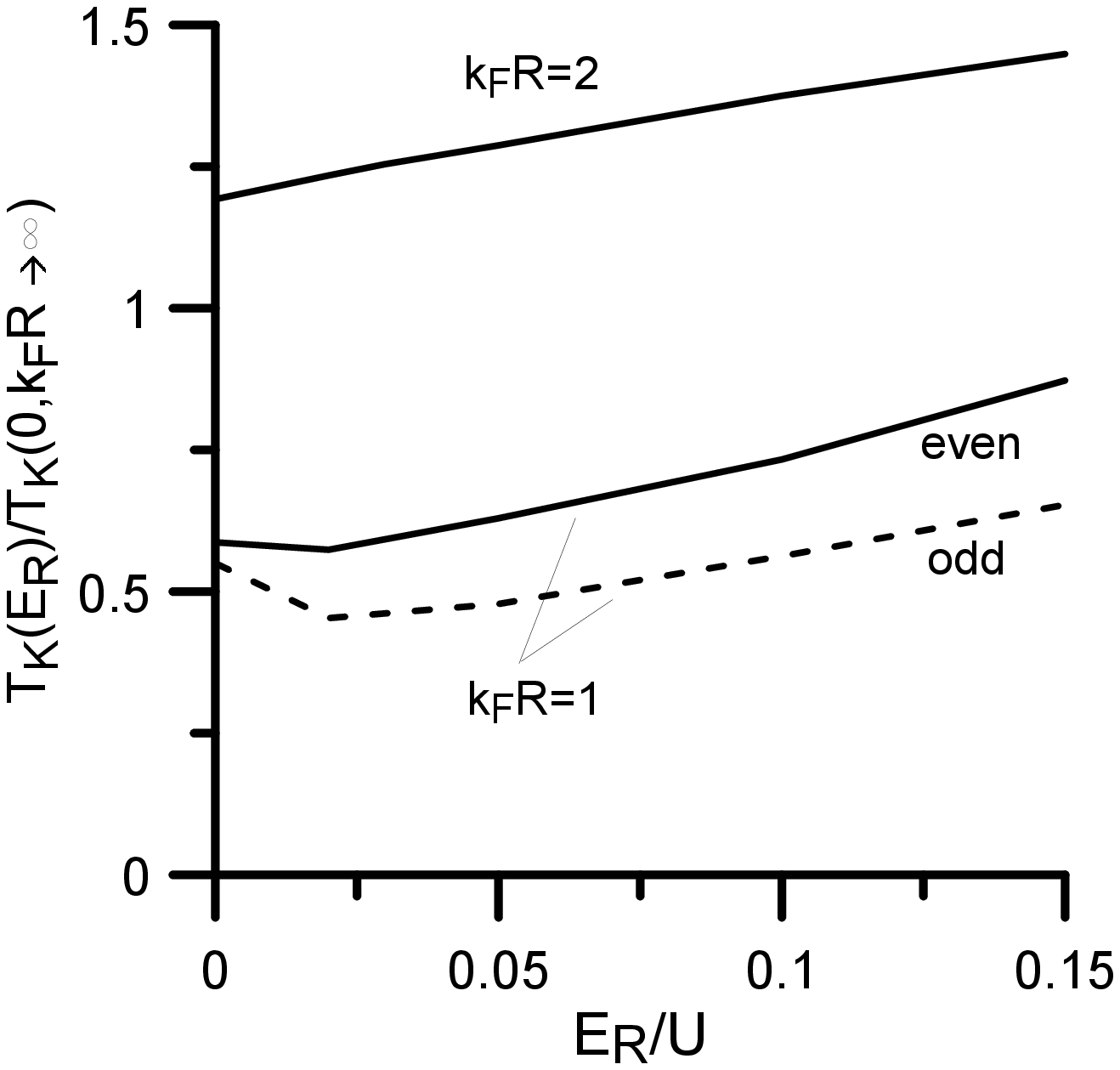}
\caption{The $E_R$ dependence of the Kondo temperature for two values $k_FR$. The data are normalized with respect to the Kondo temperature for $E_R=0$ in the limit $k_FR \to \infty$.}
\label{fig4}
\end{figure}

\begin{thebibliography}{*}
\bibitem{focus} G. E. W. Bauer  and L. W. Molenkamp (eds), {\it Focus on spintronics in reduced dimensions}, New J. Phys. {\bf 9} (2007).
\bibitem{fabian} J. Fabian, A. Matos-Abiague, C. Erstler, P. Stano, and I. \u{Z}utic, Acta Physica Slovaca {\bf 57}, 565 (2007).
\bibitem{awschalom} D. D. Awschalom, D. Loss, and N. Samarth (eds), {\it Semiconductor spintronics and quantum computing} (Springer-Verlag Berlin Heidelberg 2002).
\bibitem{winkler} R. Winkler, {\it Spin-orbit coupling effects in two-dimensional electron and hole systems} (Springer-Verlag Berlin Heidelberg 2003).
\bibitem{hewson} A. C. Hewson, {\it The Kondo problem to heavy fermions} (Cambridge Univ. Press 1993).
\bibitem{gordon} D. Goldhaber-Gordon, H. Shtrikman, D. Mahalu, D. Abusch-Magder, U. Meirav, and M. A. Kastner, Nature \textbf{391}, 156 (1998);
S. M. Cronenwett, T. H. Oosterkamp, and L. P. Kouwenhoven, Science \textbf{281}, 540 (1998).
\bibitem{li} J. Li, W. D. Schneider, R. Berndt, and B. Delley, Phys. Rev. Lett. \textbf{80}, 2893 (1998);
V. Madhavan, W. Chen, T. Jamneala, M. F. Crommie, and N. S. Wingreen, Science \textbf{280}, 567 (1998).
\bibitem{rashba} E. I. Rashba and V. I. Sheka, Sov. Phys. Solid State \textbf{3}, 1257 (1961).
\bibitem{brar} V. W. Brar et al., Nature Phys. \textbf{7}, 43 (2011).
\bibitem{malecki} J. Malecki, J. Stat. Phys. \textbf{129}, 741 (2007).
\bibitem{zarea} M. Zarea, S. E. Ulloa, and N. Sandler, Phys. Rev. Lett. \textbf{108}, 046601 (2012).
\bibitem{zitko} R. \u{Z}itko and J. Bon\u{c}a, Phys. Rev. B \textbf{84}, 193411 (2011).
\bibitem{krishna} H. R. Krishna-murthy, J. W. Wilkins, and K. G. Wilson, Phys. Rev. B \textbf{21}, 1003 (1980); C. Jayaprakash, H. R. Krishna-murthy, and J. W. Wilkins, J. Appl. Phys. \textbf{53}, 2142 (1982).
\bibitem{fye} R. M. Fye, J. E. Hirsch, and D. J. Scalapino, Phys. Rev. B \textbf{35}, 4901 (1987).
\bibitem{santoro} G. E. Santoro and G. F. Giuliani, Phys. Rev. B \textbf{49}, 6746 (1994).
\bibitem{ivanov} T. I. Ivanov, Phys. Rev. B \textbf{62}, 12 577 (2000).
\bibitem{mross} D. F. Mross and H. Johannesson, Phys. Rev. B \textbf{80}, 155302 (2009); H. Johannesson, D. F. Mross, and Erik Eriksson, Mod. Phys. Lett. B, \textbf{25}, 1083 (2011).
\bibitem{martin} A. Martin-Rodero, F. Flores, M. Baldo, and R. Pucci, Solid State Commun. \textbf{44}, 911 (1982); D. Meyer, T. Wegner, M. Potthoff, and W. Nolting, Physics B \textbf{270}, 225 (1999).
\bibitem{smetka1} In deriving Eq. (\ref{eq6}) we have used the well known expansion $e^{i\alpha\sin x}=\sum^{\infty}_{m=-\infty} e^{im x} J_m(\alpha)$.
\bibitem{smetka2}In order to obtain this result we have used the relations $\sum^{\infty}_{m=-\infty}[1 \pm (-1)^m] J^2_m(z)=1 \pm J_0(2z)$.
\bibitem{potthoff} M. Potthoff, T. Wegner, and W. Nolting, Phys. Rev. B {\bf 55}, 16 132 (1997).
\bibitem{janke} E. Janke, F. Emde, and F. L\"{o}sch, {\it Tafeln H\"{o}herer Funktionen}, Ch. XIII (B. G. Teubner Verlagsgesellschaft, Stuttgart 1960).

\end{thebibliography}
\end{document}